\documentclass{ifacconf}

\usepackage{natbib}        % required for bibliography

\usepackage{graphicx}      % include this line if your document contains figures
\usepackage{amsmath,amssymb,bm}
\usepackage{xcolor}
\usepackage{fp, tikz, pgfplots, tabularx}
\usetikzlibrary{arrows}
\usetikzlibrary{shapes}
\usetikzlibrary{patterns}
\usetikzlibrary{decorations.pathmorphing}
\usetikzlibrary{decorations.pathreplacing}
\usetikzlibrary{decorations.shapes}
\usetikzlibrary{decorations.text}
\usetikzlibrary{shapes.misc}
\usetikzlibrary{decorations.markings}
\usetikzlibrary{arrows.meta}
\usetikzlibrary{backgrounds}
\usepgfplotslibrary{fillbetween}

\newtheorem{ass}{Assumption}

\newtheorem{remark}{Remark}

\newcolumntype{Y}{>{\centering\arraybackslash}X}

\makeatletter
\renewcommand{\paragraph}{%
	\@startsection{paragraph}{4}%
	{\z@}{0.5ex \@plus 0.0ex \@minus .2ex}{-1em}%
	{\normalfont\normalsize\itshape}%
}
\makeatother

    \makeatletter
    \renewcommand*{\@fnsymbol}[1]{\ensuremath{\ifcase#1\or \or \dagger\or \ddagger\or
    		\mathsection\or \mathparagraph\or \|\or **\or \dagger\dagger
    		\or \ddagger\ddagger \else\@ctrerr\fi}}
    \makeatother
    
\newcommand\copyrighttext{%
	\footnotesize \textcopyright 2020 the authors. This work has been accepted to IFAC for publication under a Creative Commons Licence CC-BY-NC-ND.}
\newcommand\copyrightnotice{%
	\begin{tikzpicture}[remember picture,overlay]
	\node[anchor=south,yshift=40pt] at (current page.south) {\fbox{\parbox{\dimexpr\textwidth-\fboxsep-\fboxrule\relax}{\copyrighttext}}};
	\end{tikzpicture}%
}

\begin{document}
\begin{frontmatter}

\title{Learning Stable Nonparametric Dynamical Systems with Gaussian Process Regression
} 

\thanks[footnoteinfo]{This work has received funding from the Horizon 2020 research and innovation programme of the European Union under grant agreement n$^\circ$ 871767 of the project ReHyb: Rehabilitation based on hybrid neuroprosthesis. A. L. gratefully  acknowledges  financial  support from  the German Academic Scholarship Foundation.}

\thanks[equal]{Both authors contributed equally.}

\author[]{Wenxin Xiao\thanksref{equal}} 
\author[]{Armin Lederer\thanksref{equal}} 
\author[]{Sandra Hirche}

\address[Second]{Chair of Information-oriented Control (ITR), Department of
	Electrical and Computer Engineering, Technical University of Munich, Germany (e-mail: {wenxin.xiao, armin.lederer, hirche}@tum.de).}

\begin{abstract}                % Abstract of not more than 250 words.
Modelling real world systems involving humans such as biological processes for disease treatment or 
human behavior for robotic rehabilitation is a challenging problem because labeled training data is 
sparse and expensive, while high prediction accuracy is required from models of these
dynamical systems. Due to the high nonlinearity of problems in this area, data-driven approaches 
gain increasing attention for identifying nonparametric models. In order to increase the 
prediction performance of these models, abstract prior knowledge such as stability should be 
included in the learning approach. One of the key challenges is to ensure sufficient flexibility
of the models, which is typically limited by the usage of parametric Lyapunov functions to
guarantee stability. Therefore, 
we derive an approach to learn a nonparametric Lyapunov function based on 
Gaussian process regression from data. Furthermore, we learn a nonparametric
Gaussian process state space model from the data and show that it is capable 
of reproducing observed data exactly. We prove that stabilization 
of the nominal model based on the nonparametric control Lyapunov function
does not modify the behavior of the nominal model at training samples. 
The flexibility and efficiency of our approach is demonstrated on the benchmark 
problem of learning handwriting motions from a real world dataset, where our 
approach achieves almost exact reproduction of the training data.
\end{abstract}


\end{frontmatter}
\copyrightnotice

\setlength{\abovedisplayskip}{2.6pt}
\setlength{\belowdisplayskip}{2.6pt}
\setlength{\floatsep}{6pt}
\setlength{\textfloatsep}{0pt}
%%%%%%%%%%%%%%%%%%%%%%%%%%%%%%%%%%%%%%%%%%%%%%%%%%%%%%%%%%%%%%%%%%%%%%%%%%%%%%%%
\section{Introduction}
Identification of models for systems involving humans is a highly relevant problem in many fields
such as medicine, where dynamical systems can be used to model the progression of a disease, 
and robotic rehabilitation, where models of the human behavior can be used to maximize the 
training efficiency. Major difficulties in these modelling problems typically are a high 
nonlinearity of real world systems, the absence of first principle models and sparsity 
of the expensive data~\citep{Pentland1999}. Therefore, parametric models are generally 
not capable of representing these complex system appropriately.\looseness=-1

As a more flexible solution, data-driven approaches, which can extract necessary information 
automatically from training data, have gained increasing attention for modeling nonlinear 
systems, since they exhibit sufficient flexibility to adapt their complexity to the observed 
data and only require marginal prior knowledge. Although classical system identification 
literature has considered the problem of determining stable models, see, e.g, \citep{Lacy2003}, 
the combination of machine learning techniques and control theory has led to a variety 
of new approaches recently. A common method is to adapt standard machine learning
approaches using Lyapunov stability constraints during model parameter optimization. Using 
a quadratic Lyapuonv function, this method has been applied to Gaussian mixture models 
in the stable estimator of dynamical systems approach proposed by \cite{Khansari-Zadeh2011}, 
and is further improved in \citep{Figueroa2018} by employing additional prior distributions, which 
ensure physical consistency. The constrained optimization method has also been used in 
combination with neural networks \citep{Neumann2013}, where the flexibility of the model
can be improved by learning the Lyapunov function with a separate neural network \citep{Lemme2014}.
Since this constrained training approach can have negative effects on the learning performance, 
it has been proposed to learn a possibly unstable nominal model and a control Lyapunov function (CLF)
separately, such that a virtual control can be determined based on the CLF to stabilize the nominal 
model \citep{MohammadKhansari-Zadeh2014}. This approach has been pursued with different 
Lyapunov functions, such as the weighted sum of asymmetric quadratic functions \citep{MohammadKhansari-Zadeh2014}
and sums of squares \citep{Umlauft2017a}. Furthermore it has been extended to achieve risk-sensitive 
behavior by considering the model uncertainty due to sparsity of data \citep{Pohler2019}.

Although the existing methods ensure stable 
trajectories and achieve low reproduction errors on many practical examples, there 
are no guarantees on the achievable expressiveness using a certain model. As this issue 
arises mainly due to the use of parametric Lyapunov functions, we develop a novel, 
nonparametric Lyapunov function which can be learned from data using Gaussian 
process regression. We employ a Gaussian process state space model (GP-SSM) 
as nominal model, and show that it can learn dynamical systems accurately on 
training data. By stabilizing the GP-SSM based on the nonparametric control Lyapunov 
function, we prove that the resulting model is capable of reproducing observed 
data exactly, while being globally asymptotically stable. The flexibility 
of the approach is demonstrated in learning dynamical systems from a real 
world dataset, and compared to existing methods.\looseness=-1

The remainder of this paper is organized as follows. In Section~\ref{sec:prob}, we describe 
the considered problem. Section~\ref{sec:sta} explains our approach to learn a stable dynamical
system which reproduces observed data. The method is compared to existing approaches on 
real world data in Section~\ref{sec:eva}.\looseness=-1

\section{Problem Statement}
\label{sec:prob}

Consider a nonlinear, discrete-time dynamical system\footnote{
\textbf{Notation:} Lower/upper case bold symbols denote vectors/matrices, 
respectively, $\bm{I}_n$ the $n\times n$ identity matrix,  
$\mathbb{R}_+$ all positive real numbers, $\|\cdot\|$ the Euclidean norm and
$E[\cdot]$ the expectation operator% and $\bm{\kappa}^{-1}$ the inverse convariance matrix
.\looseness=-1}
\begin{align}
\label{eq:truesys}
    \bm{x}_{k+1}=\bm{f}(\bm{x}_k)
\end{align}
which is asymptotically stable on the continuous valued state space 
$\mathbb{X}\subset\mathbb{R}^d$. Furthermore, assume that the function 
$\bm{f}(\cdot)$ is unknown and that consecutive measurements 
of the states are taken such that we obtain a training data set 
$\mathbb{D}=\{(\bm{x}_k^{(m)},\bm{x}_{k+1}^{(m)})\}_{m=1}^M$ with $M\in\mathbb{N}$
data pairs. We will make use of the following assumption.
\begin{ass}
	The function $\bm{f}(\cdot)$ defines an asymptotically stable 
	system \eqref{eq:truesys} on the compact set~$\mathbb{X}\subset\mathbb{R}^d$.%\looseness=-1
\end{ass}
We want to estimate a model based on the observed data, which 
exhibits the posed assumptions on stability. Therefore, 
the goal is to derive a stable model of the unknown dynamical system, 
which maximizes the accuracy of the reproduced training trajectories
by reproducing the  observed data $(\bm{x}_k^{(m)},\bm{x}_{k+1}^{(m)})$
exactly.

\section{Stabilization of Gaussian Process State Space Models}
\label{sec:sta}

For learning stable dynamical systems capable of reproducing 
observations, we follow the control Lyapunov function approach proposed
in \citep{Umlauft2017a}. For this virtual stabilization method, we separately learn a nominal
system model $\bm{\mu}:\mathbb{X}\rightarrow\mathbb{R}^d$  and a control Lyapunov function
$V:\mathbb{X}\rightarrow\mathbb{R}_+$ from the training data. For a prediction, we determine the 
optimal, stabilizing virtual control $\bm{u}^*(\bm{x})$ for the nominal model $\bm{\mu}(\cdot)$ based on 
the control Lyapunov function $V(\cdot)$, which minimally modifies the nominal model, and define the 
stable model as
\begin{equation}
	\hat{\bm{f}}(\bm{x})=\bm{\mu}(\bm{x})+\bm{u}^*(\bm{x}).
     \label{overall}
\end{equation}
Since we consider scenarios with sparse data, we employ Gaussian process (GP) regression,
whose implicit bias-variance trade-off avoids overfitting and hence, provides high 
prediction accuracy with few training samples. We consider deterministic systems and 
therefore, we use noise-free Gaussian process state space models as nominal model in contrast 
to the approach proposed in \citep{Umlauft2017a}. We show that the noise-free GP-SSMs are 
capable of reproducing the training data exactly under weak assumptions
in Section~\ref{subsec:GPSSM}.
In Section~\ref{subsec:GPCLV} we propose a novel 
method to learn a nonparametric 
control Lyapunov function from training data based on Gaussian process regression,
which is guaranteed to converge along the training data. 
Finally, we show that a stabilizing control can be obtained via a constrained 
optimization and equals zero for all training data in Section~\ref{subsec:stabil}. 
Therefore, we obtain an asymptotically stable model \eqref{overall}, which is capable 
of reproducing observed data exactly.\looseness=-1

\subsection{Gaussian Process State Space Models}
\label{subsec:GPSSM}
Gaussian processes are a powerful machine learning tool for approximating nonlinear 
functions \citep{Rasmussen2006}. A GP is a stochastic process on the continuous input 
domain $\mathbb{X}$ such that each finite subset $\{\bm{x}_1,\ldots,\bm{x}_N\}\subset\mathbb{X}$ 
is assigned a joint Gaussian distribution. This view is equal to a consideration as distribution 
over functions, which is typically expressed through 
\begin{equation}
    f(\bm{x}) \sim \mathcal {GP}(m(\bm{x}),k(\bm{x},\bm{x}'))
\end{equation}
with prior mean and covariance function 
\begin{align}
     m(\bm{x})&=E[f(\bm{x})]\\
     k(\bm{x},\bm{x}')&=E[(f(\bm{x})-m(\bm{x}))(f(\bm{x}')-m(\bm{x}'))]. 
\end{align}
A GP is completely specified by its mean function $m(\cdot)$ and covariance 
kernel $k(\cdot,\cdot)$. 
The mean function allows to include prior knowledge in the form of approximate 
or parametric models. While such models exist for some applications, we do 
not assume their availability in the following and set the prior mean function 
to \mbox{$m(\bm{x})=0$} without loss of generality. The covariance kernel 
$k(\cdot,\cdot)$ is used to encode more abstract prior knowledge such as 
information about the smoothness of the regressed function and determines 
which functions can be approximated properly with a Gaussian process. Probably
the most commonly used kernel is the squared exponential (SE) kernel with 
automatic relevance determination
\begin{equation}
k(\boldsymbol{x},\boldsymbol{x'})=\sigma_f^{2}\mathrm{exp}\bigg(-\frac{1}{2}\sum\limits_{i=1}^d\bigg(\frac{x_{i}-x'_{i}}{l_i}\bigg)^{2}\bigg)
\label{kernel},
\end{equation}
where $\sigma_f^{2} \in\mathbb{R}_+$ is the signal variance and 
$l_i\in\mathbb{R}_+,\forall i=1,\ldots,d$ are the length-scale parameters. 
These variables are concatenated in a hyperparameter vector 
$\bm{\psi}=[l_1 \ldots  l_d\ \sigma_f]^T$. The squared exponential kernel is 
a universal kernel in the sense of \citep{Steinwart2001} which means 
that it allows to approximate continuous functions arbitrarily well. 
Therefore, Gaussian process regression with this kernel is 
capable of learning many typical dynamics.

We employ $d$ independent GPs to model a dynamical system with $d$-dimensional state space, 
such that the $i$-th component is denoted by 
\begin{equation}
\label{eq:priorGP}
    f_{i}(\bm{x})\sim\mathcal {GP}(0,k_i(\bm{x,x}')).
\end{equation}
Predictions with this model can be calculated by conditioning the prior GPs \eqref{eq:priorGP} 
on the given training set $\mathbb{D} = \{\bm{x}_k^{(n)},\bm{x}_{k+1}^{(n)}\}_{i=1}^N$. 
The conditional expectation $\mu(\cdot)$ can be 
calculated analytically using linear algebra. For this reason, we define target vectors
\begin{align}
    \bm{y}_i=[x_{k+1,i}^{(1)}\ \ldots\ x_{k+1,i}^{(N)}].
    \label{eq:y_i}
\end{align}
Then, the predictive mean is given by
\begin{align}
    \mu_i(\bm{x})=\bm{k}_{i}(\bm{x})\bm{K}_i^{-1}\bm{y}_i\label{prediction}
\end{align}
with $K_{i,n,m}=k_i(\bm{x}_k^{(n)},\bm{x}_k^{(m)})$ and $k_{i,n}=k_i(\bm{x},\bm{x}_k^{(n)})$. 
\begin{remark}
The matrix inverse in \eqref{prediction} theoretically always exists for the 
squared exponential kernel if there is no repeated entry in the input data, i.e., 
$\bm{x}_k^{(n)}\!\neq\! \bm{x}_k^{(m)}$, $\forall n\!\neq\! m$, due to the fact that this kernel 
is universal \citep{Steinwart2001}. However, a small regularizer, typically called 
observation noise variance 
$\sigma_{\mathrm{on}}^2$, can be added on the diagonals of $\bm{K}_i$ in order to avoid 
numerically ill-conditioned inversions. This regularizer has typically a small effect on
the prediction since the resulting mean squared prediction error is smaller than the 
noise variance \citep{Rasmussen2006}.\looseness=-1
\end{remark}

The hyperparameters $\bm{\psi}_i$ of the Gaussian processes $f_i(\cdot)$ can be obtained by independently 
maximizing the log-likelihood
\begin{align}
\log p(\boldsymbol{y}_i|\boldsymbol{X}_i)\!=\!&-\!\frac{1}{2}\boldsymbol{y}_i^{T}\!\boldsymbol{K}_i^{-\!1}\!\boldsymbol{y}_i\!-\!\frac{1}{2}\!\log \det(\!\boldsymbol{K}_i\!)\!-\!\frac{N}{2}\!\log 2\pi\!,\!
\label{eq:loglik}
\end{align}
where $\bm{X}=[\bm{x}_k^{(1)}\ \ldots\ \bm{x}_k^{(N)}]$ denotes the input training data matrix.
This optimization problem is typically solved using gradient based approaches \citep{Rasmussen2006}, 
even though it is generally non-convex.

We use the posterior mean function $\bm{\mu}(\cdot)$ defined through~\eqref{prediction} to 
define a nominal dynamical model
\begin{align}
    \label{eq:nomsys}
    \bm{x}_{k+1}=\bm{\mu}(\bm{x}_k)
\end{align}
which is generally not asymptotically stable. However, training samples
are reproduced exactly such that we obtain the following result.
\begin{lem}
\label{th:GPSSM}    
    Consider a training data set $\mathbb{D}\!=\!\{\!(\!\bm{x}_k^{(m)}\!\!\!,\bm{x}_{k+1}^{(m)}\!)\!\}_{m\!=\!1}^M$ 
    generated by an unknown dynamical system~\eqref{eq:truesys}, which has a stable 
    equilibrium at the origin. Furthermore, assume that
    the training data set is augmented by adding the pair~$(\bm{0},\bm{0})$. 
    Then, a Gaussian process state space model trained with this training data set
    reproduces the training data exactly and has an equilibrium at the origin.
\end{lem}
\begin{pf}
    Performing the prediction for all training inputs~$\bm{x}_k^{(m)}$ jointly
    yields
    \begin{align*}
        \mu_i(\bm{X}_k)=\bm{K}_i\bm{K}_i^{-1}\bm{y}_i=\bm{y}_i,
        %\label{mu}
    \end{align*}
    where
    $\mu_i(\bm{X}_k)=[\mu_i(\bm{x}_k^{(1)})\ \ldots\ \mu_i(\bm{x}_{k}^{(M)})\ \mu_i(\bm{0})]$
    and the inverse is well defined due to the fact that we consider a deterministic
    function $\bm{f}(\cdot)$ such that $\bm{x}_{k+1}^{(m)}\neq\bm{x}_{k+1}^{(m')}$, $\forall m\neq m'$.
    Furthermore, we have the identity
    \begin{align*}
        [\bm{y}_1\ \ldots\ \bm{y}_d]^T=[\bm{x}_{k+1}^{(1)}\ \ldots \ \bm{x}_{k+1}^{(M)}\ \bm{0}]
    \end{align*}
    due to the definition of the data set $\mathbb{D}$. 
    Therefore, the training data is reproduced exactly by the nominal system 
    \eqref{eq:nomsys}.  
    Finally, the equilibrium at the origin follows from the additional 
    training pair $(\bm{0},\bm{0})$ due to \citep{Umlauft2018}.\looseness=-1
\end{pf}
The exact reproduction of data regardless of their complexity 
is a major advantage of the nonparametric GP modeling approach. However, this reproduction
is only possible, if the training data can be considered noise-free, which is exploited 
in the proof as the property $\bm{x}_{k+1}^{(m)}\!\neq\!\bm{x}_{k+1}^{(m')}$. In applications
with few training data such as medical applications or human-robot interaction, this condition is 
typically satisfied due to the sparsity of the data. Therefore, it 
is not a severe restriction.\looseness=-1

\begin{remark}
Since we only focus on deterministic systems in our approach, the variance of the next state $\bm{x}_{k+1}$ is not of primary interest in this paper. However, it could be used to determine regions 
of the state space $\mathbb{X}$, which require more training data in order to provide a good 
model of the dynamical system.\looseness=-1
\end{remark}

\subsection{Learning Nonparametric Control Lyapunov Functions}
\label{subsec:GPCLV}
Although exact reproduction of the data is possible using GP-SSMs, 
this does not imply that the stabilized system \eqref{overall}
also exhibits superior reproduction performance. This is due to the fact that an insufficiently flexible
parameterization of the control Lyapunov function $V(\cdot)$ might not allow the decrease of
$V(\cdot)$ along all training samples. However, the required flexibility is
difficult to determine a priori with parametric 
functions such as sums of squares or weighted sum of asymmetric quadratic functions \citep{Umlauft2017a}.
Therefore, we propose to learn a 
control Lyapunov function from data based on Gaussian process regression
to exploit the flexibility of a fully nonparametric approach. Since we do not have any target values 
for the supervised learning, we cannot directly apply the GP regression approach. 
Therefore, we approximate the infinite horizon cost $\tilde{V}_{\infty}(\bm{x})=\sum_{k=1}^{\infty}l(\bm{f}^k(\bm{x}))$,
where $\bm{f}^k(\cdot)$ denotes the $k$-times application of the 
dynamics $\bm{f}(\cdot)$ and $l:\mathbb{R}^d\rightarrow\mathbb{R}_+$ is a chosen stage cost, 
by transforming the Bellman equation at training points into a regression problem 
as proposed in \citep{Lederer2019b}. 
This is formalized in the following lemma.
\looseness=-1
\begin{lem}\label{th:CLF}
Consider the approximate infinite horizon cost
\begin{align}
    V_{\infty}(\bm{x})=\bm{\lambda}^T(\bm{k}(\bm{X}_k,\bm{x})-\bm{k}(\bm{X}_{k+1},\bm{x})),
    \label{eq:Vinfty}
\end{align}
with positive definite stage cost $l:\mathbb{R}^d\rightarrow\mathbb{R}_+$, 
training points $\bm{X}_k=[\bm{x}_k^{(1)}\ \ldots\ \bm{x}_k^{(M)}]$, 
$\bm{X}_{k+1}=[\bm{x}_{k+1}^{(1)}\ \ldots\ \bm{x}_{k+1}^{(M)}]$ and\looseness=-1
\begin{align}
    \bm{k}(\bm{X},\bm{x})&= \left[k(\bm{x}^{(1)},\bm{x})\ \ldots\ k(\bm{x}^{(M)},\bm{x})\right]^T\\
    \bm{\lambda}&=\bm{\kappa}^{-1}[l(\bm{x}_{k+1}^{(1)})\ \ldots\ l(\bm{x}_{k+1}^{(M)})]^T,
    \label{eq:lambda}
\end{align}
where the elements of the invertible matrix $\bm{\kappa}\in\mathbb{R}^{M\times M}$ are 
defined using the squared exponential kernel $k(\cdot,\cdot)$ as
\begin{align}
    \kappa_{mn}&= k\left(\bm{x}_k^{(m)},\bm{x}_k^{(n)}\right) - k\left(\bm{x}_{k+1}^{(m)},\bm{x}_k^{(n)}\right)\nonumber\\
    &-k\left(\bm{x}_k^{(m)},\bm{x}_{k+1}^{(n)}\right) + k\left(\bm{x}_{k+1}^{(m)},\bm{x}_{k+1}^{(n)}\right).
    \label{eq:kappamn}
\end{align}
Then, the control Lyapunov  function 
\begin{align}
    V(\bm{x})=l(\bm{x})+\max\{0,V_{\infty}(\bm{x})+V_{\infty}(\bm{0})\}
    \label{eq:V}
\end{align}
is positive definite and decreasing along the training data, i.e., 
$V(\bm{x}_k^{(m)})\!\geq\! V(\bm{x}_{k+1}^{(m)})$, $\forall m\!=\!1,\ldots,M$.
\end{lem}
\begin{pf}
Since $l(\cdot)$ is a positive definite function, $V(\cdot)$ is positive 
due to its definition \eqref{eq:V}. Hence, it remains to show the 
decrease along the training data. For this reason, we first consider the 
exact infinite horizon cost function $\tilde{V}_{\infty}(\cdot)$, which 
satisfies the Bellman equation
\begin{align*}
    \tilde{V}_{\infty}(\bm{x})-\tilde{V}_{\infty}(\bm{f}(\bm{x}))=l(\bm{f}(\bm{x})).
\end{align*}
Due to \citep{Lederer2019b}, a function satisfying this equation on a 
finite set of pairs $(\bm{x},\bm{f}(\bm{x}))$ can be obtained through noiseless
GP regression with the kernel 
\begin{align*}
    \tilde{k}(\bm{x},\bm{x}')\!=\!k(\bm{x},\!\bm{x}')\!-\!k(\!\bm{f}(\bm{x}),\!\bm{x}')\!-\!k(\bm{x},\!\bm{f}(\bm{x}')\!)\!+\!k(\!\bm{f}(\bm{x}),\!\bm{f}(\bm{x}')\!)
\end{align*}
and output training data $\bm{y}\!=\![l(\bm{x}_{k+1}^{(1)})\ \ldots\ l(\bm{x}_{k+1}^{(M)})]^T\!$,
where invertibility
of the matrix $\bm{\kappa}$ defined through \eqref{eq:kappamn} is guaranteed due to the 
usage of the universal squared exponential kernel. This follows directly from a representation
of $\tilde{V}_{\infty}(\cdot)$ in the feature space associated with the kernel $k(\cdot,\cdot)$ 
and the linearity of this representation. Substituting the obtained regression result 
into the original feature space representation of $\tilde{V}_{\infty}(\cdot)$ directly 
yields \eqref{eq:Vinfty}. Since 
the approximate infinite horizon cost \eqref{eq:Vinfty} is guaranteed to satisfy the 
Bellman equation on the training
pairs, the approximated cost $l(\bm{x})+V_{\infty}(\bm{x})$ is decreasing along 
the training data. Since this property is shift invariant, i.e., adding a constant to 
\eqref{eq:Vinfty} does not change the decrease along training data, \eqref{eq:Vinfty} is 
not guaranteed to be positive for all $\bm{x}\neq \bm{0}$. Therefore, we enforce 
$V(\bm{0})=0$ by adding the value of \eqref{eq:Vinfty} evaluated at the origin and exclude obvious 
regression errors by setting negative values of the shifted approximate infinite horizon 
$V_{\infty}(\bm{x})+V_{\infty}(\bm{0})$ to~$0$. Since regression errors do not occur on the 
training samples, the decrease along training data is guaranteed for \eqref{eq:V} and the theorem
is proven.\looseness=-1
\end{pf}

The hyperparameters of the control Lyapunov function
$V(\cdot)$ can be obtained via the standard approach of maximizing the 
log-likelihood \eqref{eq:loglik}. However, if we assume a parameterized stage cost 
$l_{\theta}(\cdot)$, we can optimize jointly with respect to hyperparameters
$\bm{\psi}$ and cost parameters $\bm{\theta}$\looseness=-1
\begin{align}
    \min\limits_{\bm{\psi},\bm{\theta}} \log p(\bm{l}_{\theta}(\bm{X}_{k+1})|\bm{X}_{k+1},\bm{X}_{k}),
\end{align}
where the log-likelihood is given by
\begin{align}
    &\log p(\bm{l}_{\theta}(\bm{X}_{k+1})|\bm{X}_{k+1},\bm{X}_{k})=\nonumber\\
    &-\!\frac{1}{2}\bm{l}_{\theta}^T(\bm{X}_{k+1})\bm{\kappa}^{-1}\bm{l}_{\theta}(\bm{X}_{k+1})\!-\! \frac{1}{2}\log(|\bm{\kappa}|)\!-\!\frac{N}{2}\log(2\pi)
    \label{eq:hypopt2}
\end{align}
with the abbreviation $\bm{l}_{\theta}(\bm{X}_{k+1})=\big[l_{\theta}(\bm{x}_{k+1}^{(1)})\ \ldots\ l_{\theta}(\bm{x}_{k+1}^{(M)})\big]$.
This approach exhibits the advantage that the highly local approximate infinite horizon 
cost $V_{\infty}(\cdot)$, which is typically nonzero only in the proximity of training data, 
and the global parametric stage cost $l(\cdot)$ are jointly adapted to the data.\looseness=-1

\begin{remark}
While we assume GPs with squared exponential kernels in this article, all theoretical results are
directly applicable to arbitrary universal kernels \citep{Steinwart2001}.
\end{remark}
\begin{remark}
	Although the Lyapunov function $V(\cdot)$ depends on the hyperparameters $\bm{\psi}$ 
	and the stage cost parameters $\bm{\theta}$, the fundamental properties such as the 
	decreasing value along the training data are not influenced by them. Therefore, the 
	Lyapunov function $V(\cdot)$ is considered nonparametric. However, the behavior 
	away from the training data crucially depends on the hyperparameters such that 
	the hyperparameter optimization \eqref{eq:hypopt2} is an important step in obtaining
	suitable hyperparameters.
\end{remark}

\subsection{Reproductivity Preserving Stabilization}
\label{subsec:stabil}
We pursue the optimization based approach proposed in \citep{Umlauft2017a} 
to virtually stabilize the nominal system \eqref{eq:nomsys} with minimal 
modification. Within this approach we obtain the stabilizing control 
$\bm{u}(\bm{x})$ through\looseness=-1
\begin{subequations}
\begin{equation}
    \bm{u}^*(\bm{x})=\arg\mathop{\min}_{\bm{u}} \frac{1}{2}\bm{u}^T\bm{u},
    \label{minu}
\end{equation}
subject to:
\begin{align}
  \begin{split}
  V(\bm{\mu}(\bm{x})+\bm{u})< V(\bm{x}) \qquad &\forall \bm{x}\neq \bm{0}    \\
  V(\bm{\mu}(\bm{x})+\bm{u})= V(\bm{x}) \qquad &\forall \bm{x}= \bm{0},
  \end{split}
  \label{constraint}
\end{align}
\label{u}
\end{subequations}
where $V(\cdot)$ is the nonparametric Lyapunov function \eqref{eq:V}. Although these non-convex 
constraints generally prevent guarantees for the global optimality of solutions, this is not 
a problem since local minima can trivially be obtained by setting $\bm{u}^*(\bm{x})=-\bm{\mu}(\bm{x})$. 
Therefore, asymptotic stability is not affected by the non-convexity of the optimization problem
which is exploited in the following theorem.
\begin{thm}
\label{th:stab}
The model \eqref{overall} with nominal model defined through \eqref{prediction}
and stabilizing control obtained in \eqref{u} based on the nonparametric control 
Lyapunov function \eqref{eq:V} with radially unbounded stage cost $l(\cdot)$ 
is globally asymptotically stable and 
reproduces training data exactly, i.e., $\hat{\bm{f}}(\bm{x}_k^{(m)})=\bm{x}_{k+1}^{(m)}$, 
for all $m=0,\ldots,M$.
\end{thm}
\begin{pf}
The function $V(\cdot)$ is positive definite and radially unbounded since the stage cost
$l(\cdot)$ also satisfies these conditions. The optimization problem is always feasible
since $\bm{u}^*(\bm{x})=-\bm{\mu}(\bm{x})$ is a trivial solution and $\|\bm{u}\|^*$ is 
bounded since each mean function of the Gaussian process state space model is bounded, i.e., 
\begin{equation*}
    |\mu_i(\bm{x})|\leq \sigma_f^{2} \sqrt{N}||\boldsymbol{K_i}^{-1}\boldsymbol{y}_i||\quad \forall i=1,\ldots,d,
\end{equation*}
with $\bm{y}_i$ from \eqref{eq:y_i}.
Because the training set is fixed and generated by a deterministic function $\bm{f}(\cdot)$ the norm of $\bm{K}_i^{-1}\bm{y}_i$ is a finite constant. 
Hence, $V(\cdot)$ is a Lyapunov function and the system \eqref{overall} is globally asymptotically
stable. Finally, reproduction of observed training data follows from the fact that $V(\cdot)$ is decreasing 
along training data as shown in Lemma~\ref{th:CLF} and the exact reproduction of training data with the nominal model \eqref{eq:nomsys} as proven in Lemma~\ref{th:GPSSM}.\looseness=-1
\end{pf}
Although we use the trivially feasible control $\bm{u}^*(\bm{x})=-\bm{\mu}(\bm{x})$
to prove asymptotic stability, it might not lead to good local optima as starting point 
of numerical optimization. 
Therefore, we propose to choose as initial point for the numerical optimization
the closest training point in the training data set (including the origin) which 
satisfies the stability conditions. This approach results in weak convergence to 
the training data as local optima in the proximity of data are more likely to 
be found.\looseness=-1

\begin{remark}
If the stability conditions are already satified by the uncontrolled GP-SSM, the optimal control 
$\bm{u}^*$ is $\bm{0}$. Therefore, there is no need to solve the optimization \eqref{u} numerically
and computation time can be reduced by directly using the nominal model \eqref{eq:nomsys}.
\end{remark}

\section{Experimental Evaluation}
\label{sec:eva}
In order to demonstrate the flexibility of the proposed nonparametric (NP) Lyapunov function,
we compare its performance to the weighted sum of asymmetric quadratic functions
(WSAQF) $V_{\rm{WSAQF}}$ \citep{MohammadKhansari-Zadeh2014} and the sum of squares (SOS) 
Lyapunov function \citep{Umlauft2017a}. We evaluate the performance in 
learning the motions of the LASA handwriting dataset\footnote{Data set is available at https://bitbucket.org/khansari/seds} because it is a well-established benchmark for stable
nonlinear dynamical systems, which fosters comparability of the methods. The setting of our 
simulations is described in Section~\ref{subsec:expset}, while the results are presented 
in Section~\ref{subsec:res} and discussed in Section~\ref{subsec:disc}.\looseness=-1

\subsection{Experimental Setting}
\label{subsec:expset}
The LASA data set consists of $24$ handwriting shapes recorded with a tablet computer. 
For each shape $3$ to $15$ recordings of the same motion are in the data set with a single trajectory
consisting of $150$ or $250$ data points. Since some of the trajectories of a single shape 
intersect and practically exhibit a stochastic behavior, our approach is not directly applicable 
to the original data. In order to ensure comparability of the control Lyapunov functions, 
we downsample the training data
by a factor $10$ to resolve this issue and obtain sparse data. For learning the GP-SSMs we add 
a regularizer $10^{-14}$ to the diagonal of the 
kernel matrices $\bm{K}_i$ and $\bm{\kappa}$ in order to improve numerical stability of 
the matrix inversion in~\eqref{prediction} and~\eqref{eq:lambda}, respectively.\looseness=-1

\paragraph*{Control Lyapunov functions}
We compare the flexibility of three different Lyapunov functions:
\begin{itemize}
    \item The WSAQF Lyapunov function proposed in \citep{MohammadKhansari-Zadeh2014} which 
    is given by 
    \begin{equation}
    V_{\mathrm{WSAQF}}(\bm{x})\!=\!\bm{x}^T \bm{P}_0 \bm{x}\!+\!\sum_{l=1}^L \beta_l(\bm{x})(\bm{x}^T \bm{P}_l(
    \bm{x}\!-\!\bm{\epsilon}_l))^2,
    \end{equation}
    where 
    \begin{equation}
    \beta_l(\bm{x})=\begin{cases}0\quad \textrm{if $\bm{x}^T \bm{P}_l  (\bm{x}-\bm{\epsilon}_l)$\textless $0$}\\1\quad \textrm{otherwise,}
    \end{cases}
    \end{equation}
    with positive definite matrices $\bm{P}_l$, $l=0,\ldots,L$. We set $L=3$ 
    in our simulations resulting in $18$ parameters.
    \item  The SOS Lyapunov function proposed in \citep{Umlauft2017a} which is 
    defined as 
    \begin{equation}
    V_{\mathrm{SOS}}(\bm{x})=\bm{m}(\bm{x})^T\bm{P}_0\bm{m}(\bm{x}),
    \end{equation} 
    where $\bm{m}(\cdot)$ is a vector of monomials and $\bm{P}_0$ is a positive definite matrix, see 
    \citep{Papachristodoulou2005} for a detailed explanation on the SOS technique. 
    We use monomials up to degree $2$ which results in $15$ free parameters.
    \item The proposed nonparametric Lyapunov function defined in \eqref{eq:V}
    which is denoted as $V_{\mathrm{NP}}(\cdot)$ in the sequel. We employ a 
    quadratic stage cost $l(\bm{x})=\bm{x}^T\bm{P}_0\bm{x}$ with positive definite matrix 
    $\bm{P}_0$ such that the conditions
    of Theorem~\ref{th:stab} for global asymptotic stability are met.\looseness=-1
\end{itemize}
The parameters $\bm{P}_l$ of the WSAQF and SOS control Lyapunov function are 
optimized to fit the data through the minimization problem
\begin{align}
    \min\limits_{\bm{P}_l} \!\sum\limits_{m=1}^M\! \max\!\left\{\!0,V(\bm{x}_{k+1}^{(m)})\!-\!
    V(\bm{x}_{k}^{(m)})\!\right\}.
\end{align}
The positive definiteness of the matrices $\bm{P}_l$ is enforced using a 
Cholesky decomposition and constraining the eigenvalues of it to be larger 
than $0.01$ in all approaches. 

\paragraph*{Simulation of the stabilized models} 
In order to compare the flexibility in reproducing the training data exactly we simulate 
the dynamical systems stabilized with the different control Lyapunov functions starting
at the initial points of each trajectory. The optimization \eqref{u} is solved using 
an interior point algorithm where the strict inequality constraint \eqref{constraint}
is enforced through
\begin{align}
    V(\bm{\mu}(\bm{x})+\bm{u})-V(\bm{x})\leq -\rho \log(1+V(\bm{x}))
\end{align}
with $\rho=0.01$ in order to improve numerical robustness. The simulation of trajectories
is stopped, if they reach a neighborhood $\|\bm{x}_k\|\!\leq\! 10$ or exceed $1000$ steps. 
We measure the reproduction error $\Delta_{\mathrm{rep}}$ between the control 
Lyapunov functions using the total area between the training trajectory and the simulated 
trajectory. In addition to these simulations, we compare the computational efficiency of 
different approaches. For this reason we measure the average time $\bar{t}_{\mathrm{train}}$ 
it takes to fit the control Lyapunov functions to the data. Furthermore, we predict the
stabilized models on a uniformly spaced $100\!\times\! 100$ grid and compare the average computation
time $\bar{t}_{\mathrm{test}}$ for a non-trivial control $\bm{u}^*(\bm{x})\!\neq\! \bm{0}$.\looseness=-1

\subsection{Results}
\label{subsec:res}
\begin{figure}
	\pgfplotsset{width=7.5cm, compat = 1.13, 
		height = 4.5cm, grid= major, 
		legend cell align = left, ticklabel style = {font=\scriptsize},
		every axis label/.append style={font=\small},
		legend style = {font=\tiny},title style={yshift=-7pt, font = \small} }

	\begin{center}
		\includegraphics[width=0.45\textwidth]{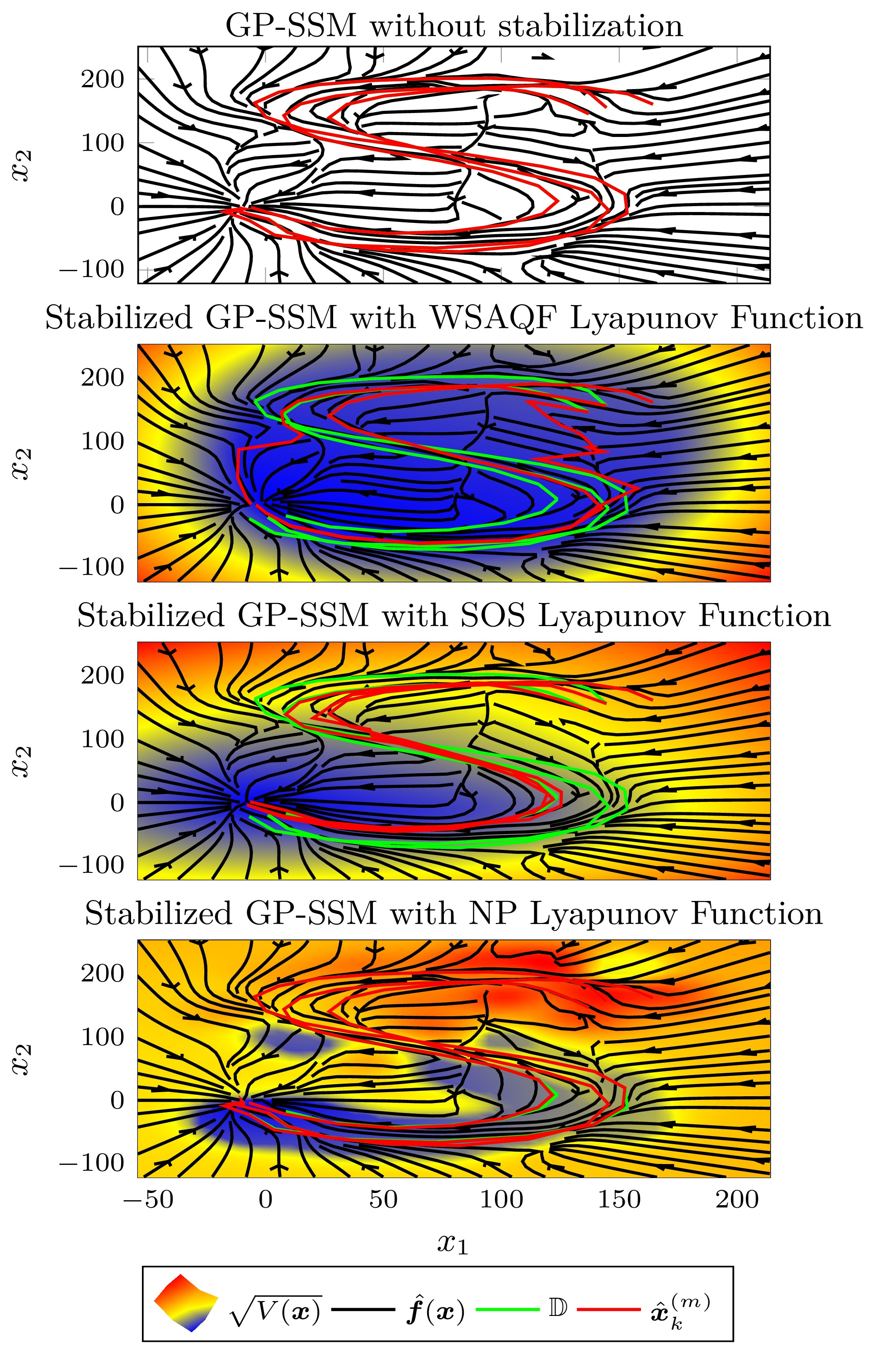}
	\end{center}
	\vspace{-0.5cm}
    \caption{GP-SSM stabilized by different control Lyapunov functions together with the 
    training data for the LASA S-shape movement.}
    \label{fig:my_label}
\end{figure}

The training data $\mathbb{D}$, the stabilized GP-SSMs $\hat{\bm{f}}(\bm{x})$ 
and the simulated 
trajectories $\hat{\bm{x}}_k^{(m)}$ for the S-shape of the LASA dataset are shown in 
Fig.~\ref{fig:my_label}. The square root of the control Lyapunov functions 
$V(\bm{x})$ are visualized by colormaps with red denoting highest and dark blue 
lowest values. In addition to the stabilized models, the GP-SSM without
stabilization is depicted which reproduces the training trajectories exactly. 
The quantitative results regarding computation times 
and reproduction errors for the S-shape as well as the whole data set 
are shown in Table~\ref{tab:perf}. It can be clearly seen that the nonparametric 
Lyapunov function provides a lower reproduction error and allows even a faster 
optimization, while it takes significantly more time to train.\looseness=-1

\setlength{\textfloatsep}{6pt}
              
\begin{table}[t]
	\vspace{1.5mm}
	\centering
	\begin{tabularx}{\columnwidth}{p{2.4cm}|Y Y Y Y}
		$V(\bm{x}_k)$& $\Delta_{\mathrm{rep}}$ S shape & $\Delta_{\mathrm{rep}}$ all 24
		& $\bar{t}_{\mathrm{train}}$ & $\bar{t}_{\mathrm{test}}$\\ \hline
		WSAQF & 14276 & 2377 & 0.7468$s$ & 0.0101$s$\\ 
		SOS & 6107 & 1819.8 & \textbf{0.4655}$s$ & 0.0083$s$\\ 
		Our method "NP" & \textbf{280.99} & \textbf{415.6} & 2.9536$s$ & \textbf{0.0056}$s$\\ 
	\end{tabularx}
	
	\caption{Reproduction error and average computation times for Lyapunov function training and stabilizing 
	control computation.}
	\label{tab:perf} 
	\vspace{-0.25cm}
\end{table}

\subsection{Discussion}
\label{subsec:disc}
The simulations clearly show that the nonparametric control Lyapunov function in combination with 
a noise-free GP-SSM allows the precise reproduction of observed training data. 
Since existing approaches such as SOS or WSAQF are limited by the number of used parameters, 
not all training samples satisfy the stability conditions. Therefore, the stabilizing control 
$\bm{u}(\bm{x})$ computed based on the SOS and WSAQF control Lyapunov functions can cause a 
deviation from the observed trajectories. In contrast, our nonparametric approach adapts its 
flexibility to the data. Although the nonparametric Lyapunov function exhibits 
local minima, this does not cause an increasing Lyapunov function along trajectories.  
Instead, a local minimum leads to discrete-time dynamics, which can have large differences
between consecutive states, since the system must move from the local minimum to a 
state with smaller Lyapunov function within a single time step. Moreover, the nonparametric
Lyapunov function approach relies on a precise nominal model: when the nominal model is too 
imprecise such that the nominal 
trajectories deviate significantly from the training data, the approximate infinite horizon cost 
$V_{\infty}(\cdot)$ is almost zero such that the quadratic stage cost $l(\cdot)$ dominates. 
Therefore, trajectories do generally not converge to training data which would be necessary 
for a risk-sensitive behavior with awareness of the sparsity of data. 
However, this does not affect stability of the 
obtained model and could be overcome by employing the approach proposed in \citep{Pohler2019}. 
Furthermore, the dominance of the quadratic cost $l(\cdot)$ exhibits also advantages regarding the 
computation time of the optimal controls $\bm{u}^*(\bm{x})$ such that the nonparametric Lyapunov
function is the fastest on average (see Table~\ref{tab:perf}).\looseness=-1

\section{Conclusion}
\label{sec:conc}

In this paper, we develop a novel approach for learning a fully nonparametric, asymptotically 
stable model, which is capable of precisely reproducing observed data. We show that deterministic
training data can be learned exactly with GP-SSMs, and employ a nonparametric control Lyapunov
function learned from the data to stabilize the nominal GP-SSM without modifying the nominal 
model at training points. In a comparison to existing GP-SSM stabilization approaches on a real 
world dataset the superior flexibility and precision of the nonparametric control Lyapunov function 
is demonstrated. In order to extend the applicability of the approach to systems with noisy data, 
we will modify the approach in future work, such that stochastic stability conditions can be considered 
for learning the nonparametric Lyapunov function.

\bibliography{mybib}             % bib file to produce the bibliography

\begin{thebibliography}{14}
\providecommand{\natexlab}[1]{#1}
\providecommand{\url}[1]{\texttt{#1}}
\providecommand{\urlprefix}{URL }
\expandafter\ifx\csname urlstyle\endcsname\relax
  \providecommand{\doi}[1]{doi:\discretionary{}{}{}#1}\else
  \providecommand{\doi}{doi:\discretionary{}{}{}\begingroup
  \urlstyle{rm}\Url}\fi

\bibitem[{Figueroa and Billard(2018)}]{Figueroa2018}
Figueroa, N. and Billard, A. (2018).
\newblock {A Physically-Consistent Bayesian Non-Parametric Mixture Model for
  Dynamical System Learning}.
\newblock In \emph{Proceedings of the Conference on Robot Learning}, volume~87,
  927--946.

\bibitem[{Khansari-Zadeh and Billard(2011)}]{Khansari-Zadeh2011}
Khansari-Zadeh, S.M. and Billard, A. (2011).
\newblock {Learning stable nonlinear dynamical systems with Gaussian mixture
  models}.
\newblock \emph{IEEE Transactions on Robotics}, 27(5), 943--957.

\bibitem[{Lacy and Bernstein(2003)}]{Lacy2003}
Lacy, S.L. and Bernstein, D.S. (2003).
\newblock {Subspace identification with guaranteed stability using constrained
  optimization}.
\newblock \emph{IEEE Transactions on Automatic Control}, 48(7), 1259--1263.

\bibitem[{Lederer and Hirche(2019)}]{Lederer2019b}
Lederer, A. and Hirche, S. (2019).
\newblock {Local Asymptotic Stability Analysis and Region of Attraction
  Estimation with Gaussian Processes}.
\newblock In \emph{Proceedings of the IEEE Conference on Decision and Control}.

\bibitem[{Lemme et~al.(2014)Lemme, Neumann, Reinhart, and Steil}]{Lemme2014}
Lemme, A., Neumann, K., Reinhart, R., and Steil, J. (2014).
\newblock {Neural Learning of Vector Fields for Encoding Stable Dynamical
  Systems}.
\newblock \emph{Neurocomputing}, 141, 3--14.

\bibitem[{{Mohammad Khansari-Zadeh} and
  Billard(2014)}]{MohammadKhansari-Zadeh2014}
{Mohammad Khansari-Zadeh}, S. and Billard, A. (2014).
\newblock {Learning control Lyapunov function to ensure stability of dynamical
  system-based robot reaching motions}.
\newblock \emph{Robotics and Autonomous Systems}, 62(6), 752--765.

\bibitem[{Neumann et~al.(2013)Neumann, Lemme, and Steil}]{Neumann2013}
Neumann, K., Lemme, A., and Steil, J.J. (2013).
\newblock {Neural learning of stable dynamical systems based on data-driven
  Lyapunov candidates}.
\newblock In \emph{IEEE International Conference on Intelligent Robots and
  Systems}, 1216--1222. IEEE.

\bibitem[{Papachristodoulou and Prajna(2005)}]{Papachristodoulou2005}
Papachristodoulou, A. and Prajna, S. (2005).
\newblock {A Tutorial on Sum of Squares Techniques for Systems Analysis}.
\newblock In \emph{Proceedings of the American Control Conference}, 2686--2700.

\bibitem[{Pentland and Liu(1999)}]{Pentland1999}
Pentland, A. and Liu, A. (1999).
\newblock {Modeling and prediction of human behavior}.
\newblock \emph{Neural Computation}, 11(1), 229--242.

\bibitem[{P{\"{o}}hler et~al.(2019)P{\"{o}}hler, Umlauft, and
  Hirche}]{Pohler2019}
P{\"{o}}hler, L., Umlauft, J., and Hirche, S. (2019).
\newblock {Uncertainty-based Human Motion Tracking with Stable Gaussian Process
  State Space Models}.
\newblock \emph{IFAC-PapersOnLine}, 51(34), 8--14.

\bibitem[{Rasmussen and Williams(2006)}]{Rasmussen2006}
Rasmussen, C.E. and Williams, C.K.I. (2006).
\newblock \emph{{Gaussian Processes for Machine Learning}}.
\newblock The MIT Press, Cambridge, MA.

\bibitem[{Steinwart(2001)}]{Steinwart2001}
Steinwart, I. (2001).
\newblock {On the Influence of the Kernel on the Consistency of Support Vector
  Machines}.
\newblock \emph{Journal of Machine Learning Research}, 2, 67--93.

\bibitem[{Umlauft et~al.(2017)Umlauft, Lederer, and Hirche}]{Umlauft2017a}
Umlauft, J., Lederer, A., and Hirche, S. (2017).
\newblock {Learning Stable Gaussian Process State Space Models}.
\newblock In \emph{Proceedings of the American Control Conference}, 1499--1504.

\bibitem[{Umlauft et~al.(2018)Umlauft, P{\"{o}}hler, and Hirche}]{Umlauft2018}
Umlauft, J., P{\"{o}}hler, L., and Hirche, S. (2018).
\newblock {An Uncertainty-Based Control Lyapunov Approach for Control-Affine
  Systems Modeled by Gaussian Process}.
\newblock \emph{IEEE Control Systems Letters}, 2(3), 483--488.

\end{thebibliography}

\end{document}